\documentclass[pdflatex,sn-mathphys-num]{sn-jnl}

\usepackage{graphicx}%
\usepackage{multirow}%
\usepackage{amsmath,amssymb,amsfonts}%
\usepackage{amsthm}%
\usepackage{mathrsfs}%
\usepackage[title]{appendix}%
\usepackage{xcolor}%
\usepackage{textcomp}%
\usepackage{manyfoot}%
\usepackage{booktabs}%
\usepackage{algorithm}%
\usepackage{algorithmicx}%
\usepackage{algpseudocode}%
\usepackage{listings}%
\usepackage{comment}
\usepackage{url}

\begin{document}


\title[Temporal patterns of preferences]{Temporal patterns of preferences through Wikipedia editing in different languages }

\author[1]{\fnm{David André} \sur{ Villamil Carrillo}}

\author*[2]{\fnm{Yérali} \sur{Gandica}}\email{yerali.gandica@uam.es}

\affil[1]{\orgname{Valencian International University}, \orgaddress{\city{Valencia}, \country{Spain}}}


\affil[2]{\orgname{Autonomous University of Madrid}, \orgaddress{\city{Madrid}, \country{Spain}}}

\affil[2]{Google Scholar: \url{https://scholar.google.com/citations?user=gbWlX60AAAAJ&hl=fr&oi=ao}}

\abstract{Temporal editing patterns on Wikipedia provide a computational lens on
cultural dynamics across linguistic communities. We analyse a decade of
editorial activity (2001--2010) in eleven Wikipedia language editions and
apply hierarchical clustering to two representations of collective
behaviour: a static one, based on the thematic composition of each
edition, and a temporal one, based on its weekly activity cycle within
each thematic category. For the temporal representation we compare two
normalisations: one retains only the shape of the weekly cycle, and the
other combines it with thematic composition. The linguistic communities
exhibit distinct circadian and weekly editing rhythms. Some groupings are
recovered by both representations and by both activity metrics (edits
and editors), most notably Russian and Chinese, whereas others, such as
German and French, emerge only from temporal activity rhythms. Temporal information thus reveals structure that thematic composition alone does not, which points
to time as a relevant and often neglected dimension in cross-cultural
analyses of digital trace data, and in studies that use online platforms
as proxies for collective cultural behaviour.}




\maketitle

\section{Introduction}\label{sec1}

Rhythmic activity is a fundamental property of living matter. All physiological processes are temporally organised in a multiplicity of cycles that link the organism to its environment \cite{Kerkhof1985,Panda2002}. Human beings are characterised by {physiologically and socially-induced circadian rhythms. Whereas physiological circadian cycles (sleeping and waking times) are influenced by biological and environmental factors, which are nowadays similar to all cultures, socially induced cycles modulate individuals differently depending on their social environment and routines. \cite{Roenneberg2003,Roenneberg2019,Carvalho2014}} There are societies with one, two, or even three peaks of resting activities during the day. These are the times when people take a break, pause at work, join a conversation, enjoy a cup of tea, or simply rest. There are societies where you can go shopping in the evenings or on Sundays but not in the middle of the day. There are others where the opposite is true. The rhythms of societies are an important sign of the identity of each culture that connects it to its past and to its environment. In general, these rhythms define all kinds of interactions between individuals, and their study can serve many interesting purposes. For example, it can help to understand societies from biological and anthropological points of view. However, it can also serve for technological and even commercial purposes.

Cultural rhythms are not only reflected in offline behaviours, but also in the ways communities engage with digital platforms. {Technology Appropriation Theory ~\cite{silverstone1996design} holds that digital technologies such as Wikipedia are not adopted uniformly across societies: each community appropriates them and integrates them into its own values, practices and rhythms. Giddens  ~\cite{Giddens2014} takes up the question of what happens once that appropriation has occurred: in his Structuration Theory, social structures are both the medium and the outcome of the practices of individuals, so that routines, once enacted, reproduce the very temporal and normative order that makes them expected. Routinisation in time and space occupies a central place in that account, precisely the dimension our analysis sets out to measure. Together, these two perspectives support the use of Wikipedia editing data as a meaningful lens through which to explore how cultural patterns shape online participation and collective knowledge production. Not only in what communities choose to document, but in when they do so.}

More broadly, the study of temporal patterns in online platforms has become a growing area within computational social science and digital behavioural research. Studies on social media platforms such as Twitter and Facebook have shown that posting activity, content sharing, and emotional expression follow culturally specific temporal signatures \cite{Golder2011, Blondel2015}. Similarly, research on mobile phone communication records has revealed that call and messaging patterns reflect deep-rooted social structures and cultural norms \cite{Eagle2009, Aledavood2015}. These findings suggest that digital traces, regardless of the platform, can serve as reliable proxies for studying collective behaviour and cultural dynamics at scale. Wikipedia, as a collaborative knowledge platform driven entirely by voluntary contributions, offers a particularly rich setting for this type of analysis, as editorial decisions reflect not only individual preferences but also collectively negotiated cultural priorities.

Circadian digital patterns have been analysed in previous studies. For example, in terms of the more elementary day-night cycle, using data collected with a smartphone app, Aledavood $\&$ Saram\"{a}ki et al., after categorising people as ``larks" (those who wake up and go to sleep early) and ``owls" (those who stay up and wake up late), they found that there is a strong connection between the chronotypes of people and the structure of their social networks. For example, owls maintain larger personal networks and occupy more central positions \cite{Aledavood2018}. 

 {Wikipedia (WP) has been used to characterise and find universalities and differences in temporal activity patterns of editors.} Specifically, Yasseri $\&$ Kertész et al. \cite{Yasseri2012} studied similarities and differences in the temporal activity patterns of editors among 34 Wikipedia repositories in different languages; the largest with respect to the number of articles. The authors analysed the average of their complete data on one day and found a global regular pattern of circadian activity, mostly similar among all the languages except for the Spanish and Portuguese {Wikipedia}. Those data-sets showed a slight delay on their activity, which could be related to time offset due to {edits from South America.} They also found interesting week-day and week-ends characterisations, which we will compare with our analysis.

Furthermore, {using Wikipedia as a proxy for cultural interest presupposes that language editions are not interchangeable representations of a shared body of knowledge, but are genuinely distinct and culturally-situated. Hecht and Gergle \cite{Hecht2010} show that over $74\%$ of concepts across 25 Wikipedia language editions exist in only one language, and that even shared concepts are often described with substantially different content, a divergence the authors attribute primarily to cultural self-focus. In another vein, motivated by analysing similarity of interests between cultural communities, and under the assumption that the number of edits on each language edition of Wikipedia is a proxy to analyse the collective interest of the corresponding language-speaking community}, Samoilenko et al. \cite{Samoilenko2016} performed a study over the 110 largest {Wikipedia} as of July 2014, covering a period from 2005 to 2013. They created networks, where nodes are Wikipedia languages and links are weighted as the strength of shared interest between them. Shared interest was computed using the Wikipedia’s interlanguage link graph to identify articles on the same concepts in different language editions. Clustering their networks of mutual interests, the authors found that linguistic similarity, shared religion, and  demographic attraction of communities are the factors that determine the similarity of interests between language communities. 

{On the same premise that the number of edits on specific topics is an indicator of the topics that are culturally relevant to a linguistic community,} and motivated by the aim of understanding the preferences of subjects across various linguistic communities, in \cite{Gandica2018} Gandica analysed the number of pages and editors for each Wikipedia category. Here, the author used a different methodology, each Wikipedia page was categorised according to the main classification tree defined 
in the English Wikipedia and maintained by its volunteer editor 
community: Arts, Sports, Rights, Events, Philosophy, Geography, 
History, Games, Mathematics, Nature, Politics, Religion, Health, 
and Food. This study provided an initial insight into the collective cultural interests of $11$ linguistic communities: Spanish, French, Portuguese, Italian, Hungarian, German, Russian, Arabic, Japanese, Chinese, and Vietnamese. The selection of these languages was based on balancing a global perspective with the size of their respective Wikipedia repositories. The analysis revealed different patterns of cultural preferences across languages, reflecting the specific thematic priorities of each linguistic community. For example, the German Wikipedia (DE-WP) exhibited the most heterogeneous distribution of category preferences, with ``Arts" being the most edited category; whereas the Italian Wikipedia (IT-WP) displayed the most homogeneous distribution. Other notable findings include ``History" being the dominant category in the Hungarian and Japanese Wikipedias, ``Nature" being most prominent in the Russian Wikipedia, and ``Politics" leading in the Vietnamese Wikipedia. Surprisingly, the study also found that the French Wikipedia contains the highest number of pages, surpassing even the English one.

{More recently, Yoon et al. \cite{Yoon2023} have addressed cross-linguistic similarity on Wikipedia from a different angle: instead of using edit counts as a proxy for interest, they construct knowledge-structure similarity networks across 59 Wikipedia language editions from the hyperlink relationships between pages and categories, treating Wikipedia as a proxy for how contemporary societies exchange information. They find that socio-economic ties, rather than geographical proximity, best explain cross-linguistic similarity in the knowledge structures of language-usage groups. We will return to this study in our Conclusions, where we contrast these interaction-based mechanisms with the proximities revealed by our temporal analysis.}

{This raises the question of what Wikipedia offers that other large-scale digital traces do not, given that culturally specific temporal signatures have already been documented on social media platforms and in mobile communication records \cite{Golder2011, Blondel2015,Eagle2009, Aledavood2015}. Our answer rests on the categorical dimension. To the best of our knowledge, Wikipedia is the only digital platform on which activity is partitioned into a general thematic taxonomy of preferences that is simultaneously community-maintained and replicated across language editions. The thematic labels available elsewhere are of a different nature: they are either imposed uniformly by the platform, and therefore cannot register cultural variation by construction, or they are defined bottom-up but have no counterpart in other languages, and therefore admit no cross-linguistic comparison. Wikipedia's category tree is neither: it spans the whole range of human knowledge, each edition maintains it through its own community of volunteer editors, and equivalent branches exist across editions, so that the same thematic coordinate can be followed from one language to another. Each edit, moreover, carries an exact timestamp and is already labelled thematically, with no classifier interposed between the behaviour and its thematic attribution. It is these properties together that make the language × category × hour representation of this study possible.}

Previous studies have examined either temporal activity patterns without thematic differentiation \cite{Yasseri2012} or cultural preferences across Wikipedia languages using static, aggregated data  \cite{Samoilenko2016,Gandica2018}. To date, however, no study has combined the temporal and categorical dimensions simultaneously, which is essential for capturing how cultural rhythms interact with thematic preferences. Motivated by exploring the possible existence of temporal patterns characterising different linguistic communities, in this paper we address this gap by performing a fine-grained temporal analysis across thematic categories and multiple languages. Specifically, we ask: What can the joint temporal-categorical editing patterns of several Wikipedia editions reveal about the circadian rhythms of different cultures? Do static and temporal clustering yield the same cultural proximities? Can we identify linguistic communities that, despite being distant in terms of culture, religion, or geography, share similar temporal editing signatures? These are the main questions we seek to answer in this study.

We show results for the same eleven languages analysed in \cite{Gandica2018}, except for the EN-WP due to its global nature of coverage. The analysis is deliberately restricted to the period 2001–2010, corresponding to the first decade of Wikipedia's existence. This temporal scope is not arbitrary: from approximately 2010 onwards, the Wikimedia Foundation implemented systematic initiatives to reduce content gaps across language editions \cite{Bao2012,Wulczyn2016}. These interventions encouraged or directed editors to create content in underrepresented languages, thereby altering the organic relationship between editorial activity and collective cultural interest. Since our study relies on the number of edits and editors as proxies for the genuine preferences of linguistic communities, including data from periods influenced by such institutional campaigns would compromise the validity of this proxy. The pre-intervention period thus provides a cleaner observational window into the spontaneous cultural dynamics reflected in Wikipedia editing behaviour.

The temporal approach allows us to understand the rhythms of how people edit in the different categories of Wikipedia. By studying these patterns, we gain a more complete understanding of human behaviour and the cultural dynamics that influence online content editing. This knowledge can be used to optimise resources and content management strategies in digital platforms. Furthermore, this analysis is useful for cross-cultural comparative studies, providing insights into the behaviour of different cultural communities. We describe details of the data {and methodology} in the next sections. Then, we show our results followed by our discussion and general conclusions. 

\section{Data and Methodology}

\subsection {The Data-sets}
The dataset analysed in this study is a comprehensive compilation covering the first decade of Wikipedia edits across multiple languages. It spans approximately the period from 2001 to 2010, providing insights into the evolution and editorial dynamics during this foundational phase when no interventions to cover the gap between languages had yet been implemented by the Wikimedia Foundation \cite{Bao2012,Wulczyn2016}. This study focuses on edits made in eleven languages: Spanish (ES), French (FR), Portuguese (PT), Italian (IT), Hungarian (HU), German (DE), Russian (RU), Arabic (AR), Japanese (JA), Chinese (ZH), and Vietnamese (VI). {These languages were inherited from \cite{Gandica2018}, where they were selected as a trade-off between a worldwide view and the sizes of the corresponding Wikipedia editions.}

{The data were obtained {in 2016} using the Wikipedia API and encompass approximately 7.8 GB. Each entry represents} an individual edit for each of the languages studied. The dataset includes columns detailing key attributes such as the user (editor), the edit, the Unix timestamp of the edit, and internal symbols used for classification. To exclude bot-generated edits, all edits with the word ``bot" (in any combination of upper- or lower-case letters) in their usernames were filtered out. Page classifications (i.e., categories) followed the main hierarchical structure defined in the English Wikipedia and maintained by its volunteer editor community. To ensure consistency across languages, the classification process began with categories in English (EN). Equivalent terms in other languages were identified, and all pages within the corresponding categories were collected using the Petscan API \cite{Petscan}, without mapping them back to the structure of the English Wikipedia. As a result, the number of pages per category varies across languages.

{We deliberately rely on the native category system of each Wikipedia edition rather than on a language-agnostic topic classifier \cite{Johnson2021}. These categories are maintained by each community to coordinate editing rather than to provide a normalised ontology, and the resulting variation in the number of pages per category across languages (\cite{Gandica2018}) is not an artefact to be corrected but the object of our study: it reflects how each linguistic community organises its thematic interest. A language-agnostic classifier, designed to maximise uniform coverage across editions, would by construction remove this variation and hence the signal we aim to measure.}

{To minimise thematic overlap, $13$ specific categories were selected from the $25$ originally defined in the English Wikipedia category tree}. The categories analysed were Arts, Sports, Rights, Events, Philosophy, Geography, History, Games, Mathematics, Nature, Politics, Religion, and Health. Categories such as Culture, Food,  Humanities, Law, Life, Matter, People, Reference Works, Science and Technology, Society, Universe, and World were excluded to avoid topic overlap. 

All timestamps were converted to the local time zone typically associated with the capital of the primary region of each version of Wikipedia in each language, which is represented in Table \ref{tab:dates}. {The conversion used the identifiers of the IANA time zone database, so that summer time was applied wherever and whenever it was in force.} This ensured that the timestamps accurately reflected their respective local time zones, allowing consistent temporal analysis across different languages and regions. {This assignment, however, is a simplification for languages spoken across multiple time zones: for Spanish and Portuguese in particular, speakers in Latin America substantially outnumber those in Spain and Portugal. This is visible in Figure \ref{fig3}, where the ES and PT activity curves appear shifted relative to the other languages; we assess this limitation further using the timezone-independent static clustering (Section \ref{sec:static}).}

It is worth noting that a given Wikipedia page may belong to more than one of the $13$ selected categories simultaneously. In such cases, the page and its associated edits are counted independently in each category to which it belongs, preserving the full categorical signal of editorial activity. Pages that did not map to any of the 
$13$ selected categories were excluded from the analysis. The data can be found at \cite{wp-data}.

{For each language, the dataset begins with the launch date of its corresponding Wikipedia, ensuring a synchronised starting point across languages.} The resulting dataset provides a solid foundation for analysing editorial preferences and patterns of human behaviour on Wikipedia. The inclusion of Unix timestamps for each edit enables detailed temporal analyses, allowing the identification of editing trends over time. Table \ref{tab:dates} summarises the timezone, initial and final dates for each language, as well as the number of edits and unique editors.

\begin{table}[h]
\centering
\begin{tabular}{@{}llllll@{}}
\toprule
Language & \textbf{Timezone} Start Date & End Date & Editors & Edits \\
\midrule
AR & Asia/Riyadh & 2003-07-11 & 2010-03-28 & 23641 & 7674946 \\
DE & Europe/Berlin & 2001-04-02 & 2010-03-28 & 357575 & 39689683 \\
ES & Europe/Madrid & 2001-05-25 & 2010-10-23 & 2682096 & 55266094 \\
FR & Europe/Paris & 2001-08-04 & 2010-03-28 & 203042 & 58325570 \\
HU & Europe/Budapest & 2003-07-09 & 2010-11-07 & 148067 & 6666339 \\
IT & Europe/Rome & 2001-09-14 & 2010-03-28 & 97162 & 22200807 \\
JA & Asia/Tokyo & 2002-09-11 & 2010-03-28 & 126639 & 24584471 \\
PT & Europe/Lisbon & 2001-06-17 & 2011-10-25 & 1475229 & 37853332 \\
RU & Europe/Moscow & 2002-11-13 & 2010-03-28 & 75929 & 14199597 \\
VI & Asia/Ho\_Chi\_Minh & 2002-11-16 & 2010-03-28 & 12259 & 3657770 \\
ZH & Asia/Shanghai & 2002-10-31 & 2010-03-28 & 76555 & 12618302 \\
\bottomrule
\end{tabular}
\caption{{Timezones,} start and end dates for each {Wikipedia} Language edition, with corresponding number of editors and edits.}
\label{tab:dates}
\end{table}

\subsection{Methods}\label{sec13}

In the static analyses, each language edition is represented as a {13-dimensional vector}, where each dimension corresponds to the total number of edits (or editors) accumulated in one of the 13 thematic categories over the full data span. In the temporal analyses, each language edition is represented as a {2,184-dimensional vector (13 categories × 168 hours), where each dimension corresponds to the average number of edits (or editors) in a given category for a given hour of the week.}

{Each edition is normalised before clustering, so that editions of very different size contribute equally. In the static analysis, the counts of each edition are divided by their sum, so that its thirteen category values add up to one. In the temporal analyses we use two complementary normalisations of the 11 × 13 × 168 array: in the first, each category of each edition is divided by its own total over the 168 hours, so that every category profile adds up to one and only the timing of the activity remains; in the second, each edition is divided by its total activity, so that its 2,184 values add up to one and the relative weight of the categories is preserved together with the timing.}

{Hierarchical clustering with Ward's linkage and Euclidean distance was applied directly to these representations, without any dimensionality reduction. Reducing the dimensionality beforehand is not a neutral operation for the question we ask: methods such as principal component analysis retain the directions of greatest variance across editions, and greatest variance is not the same as greatest relevance, so a similarity confined to a few of the 2,184 coordinates may be discarded. We report in the Supplementary Information what happens when the data are reduced with PCA, an autoencoder, t-SNE and UMAP before clustering. The analytical methodologies and code-base are publicly available at ~\cite{url}.}

\section{Results}\label{sec2}

\subsection{Temporal evolution}

\begin{figure}[h]
\includegraphics[width=1\linewidth]{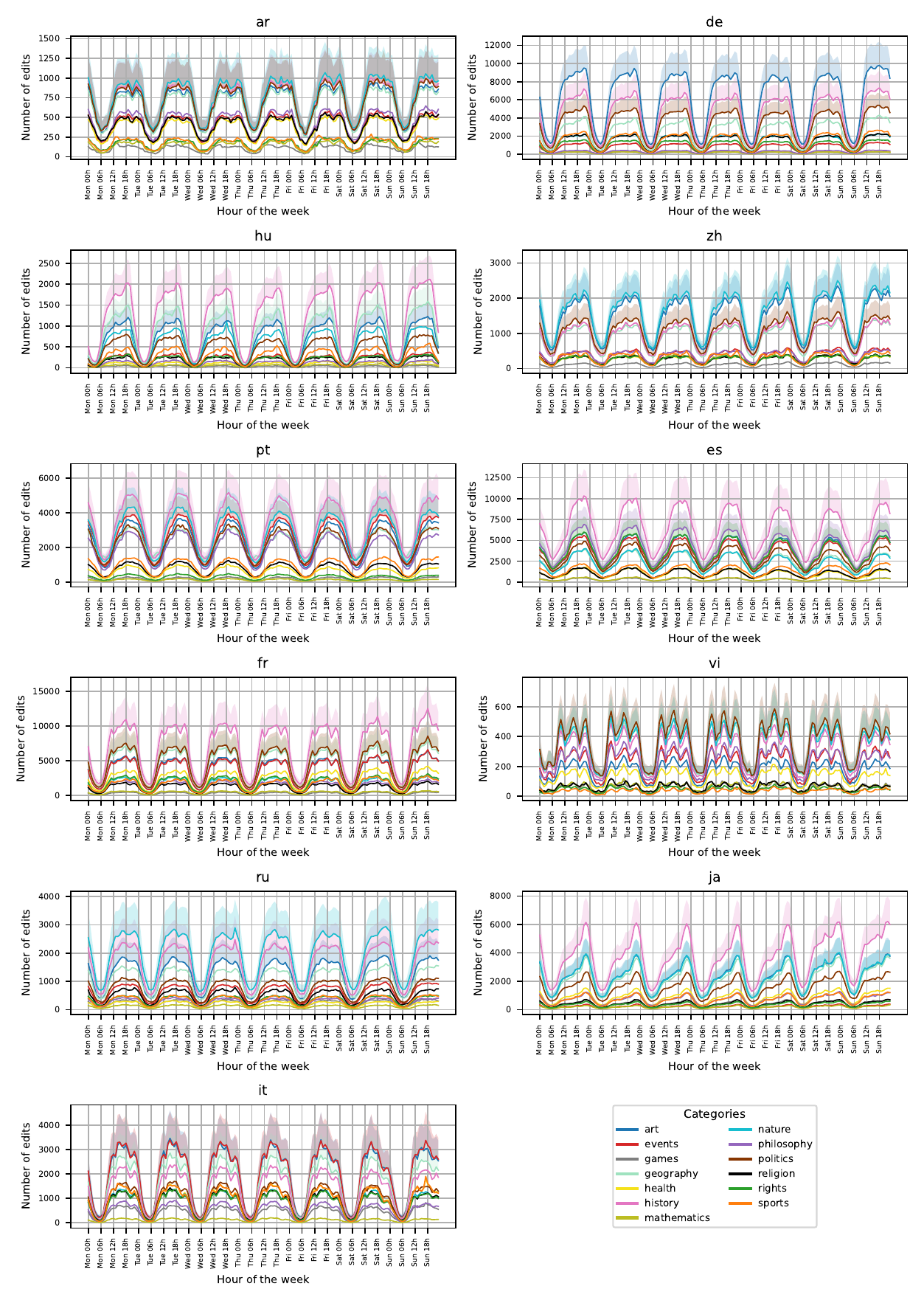}
\caption{{Average number of edits per hour across all categories for the 11 analysed languages throughout the week. The line shows the yearly mean; the shaded band shows +1 standard error of the mean (SEM) above the line, drawn only for the three most active categories in each language for readability. Times are expressed in each language's local timezone}.
\label{fig1}}
\end{figure}

\begin{figure}[h]
\includegraphics[width=1\linewidth]{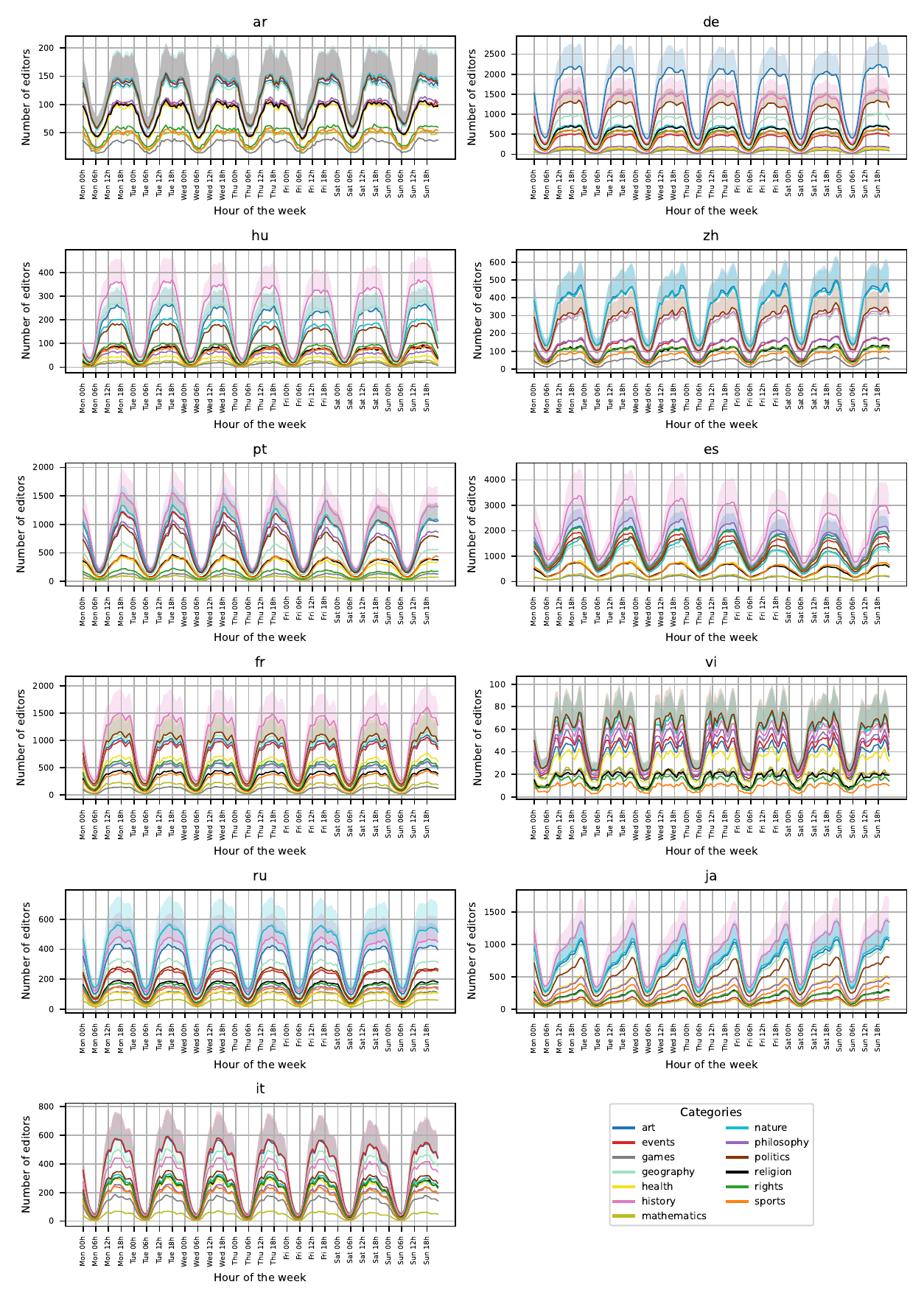}
\caption{{Average number of editors per hour across all categories for the 11 analysed languages throughout the week. The line shows the yearly mean; the shaded band shows +1 standard error of the mean (SEM) above the line, drawn only for the three most active categories in each language for readability. Times are expressed in each language's local timezone}.
\label{fig2}}
\end{figure}

{Figures \ref{fig1} and \ref{fig2} highlight notable variations across all categories within each language, beyond the expected physiological circadian patterns of activity and rest. For instance, the Vietnamese Wikipedia exhibits a distinct behaviour, with three pronounced daily activity peaks around 09:00, 15:00, and 21:00,  a pattern that contrasts with the one or two peaks typically observed in most other language editions. However, the Japanese Wikipedia also differs from the others by exhibiting a single pronounced peak at 21:00-22:00, which it maintains at the same hour throughout the week, with no distinction between weekdays and weekends; while the Italian version shows a less marked but consistent peak around {15:00–16:00} (except on Sundays). The Portuguese version displays a distinct peak between 18:00 and 19:00, present on all days of the week and more pronounced in the number of editors than in the number of edits.}

{Furthermore, ES, PT, and IT display higher activity levels on weekdays than during weekends. ES and PT from Monday to Thursday (PT mainly on the number of editors), while IT also reaches Fridays (only taking into account the number of editors as well). In contrast, FR and ZH are more active at weekends. FR is most active on Sundays, while ZH all weekend. DE shows a decrease in activity in the number of editors on Saturdays, while in the number of edits on Fridays and Saturdays.}

{In terms of preferences into categories, History appears to be the preferred category for HU, ES, FR, {PT} and JA, while Arts for DE, and Nature for RU {and ZH}. Then, there is the case of IT for which Events and Arts seem equally important. Both figures reveal additional language-specific patterns, which can be further examined in conjunction with the findings from \cite{Gandica2018}.} 

These deviations indicate that while global (physiological) circadian patterns in Wikipedia editing behaviour exist, each linguistic community exhibits its unique socially-induced temporal trends. 

\begin{figure}
    \centering
    \includegraphics[width=0.7\linewidth]{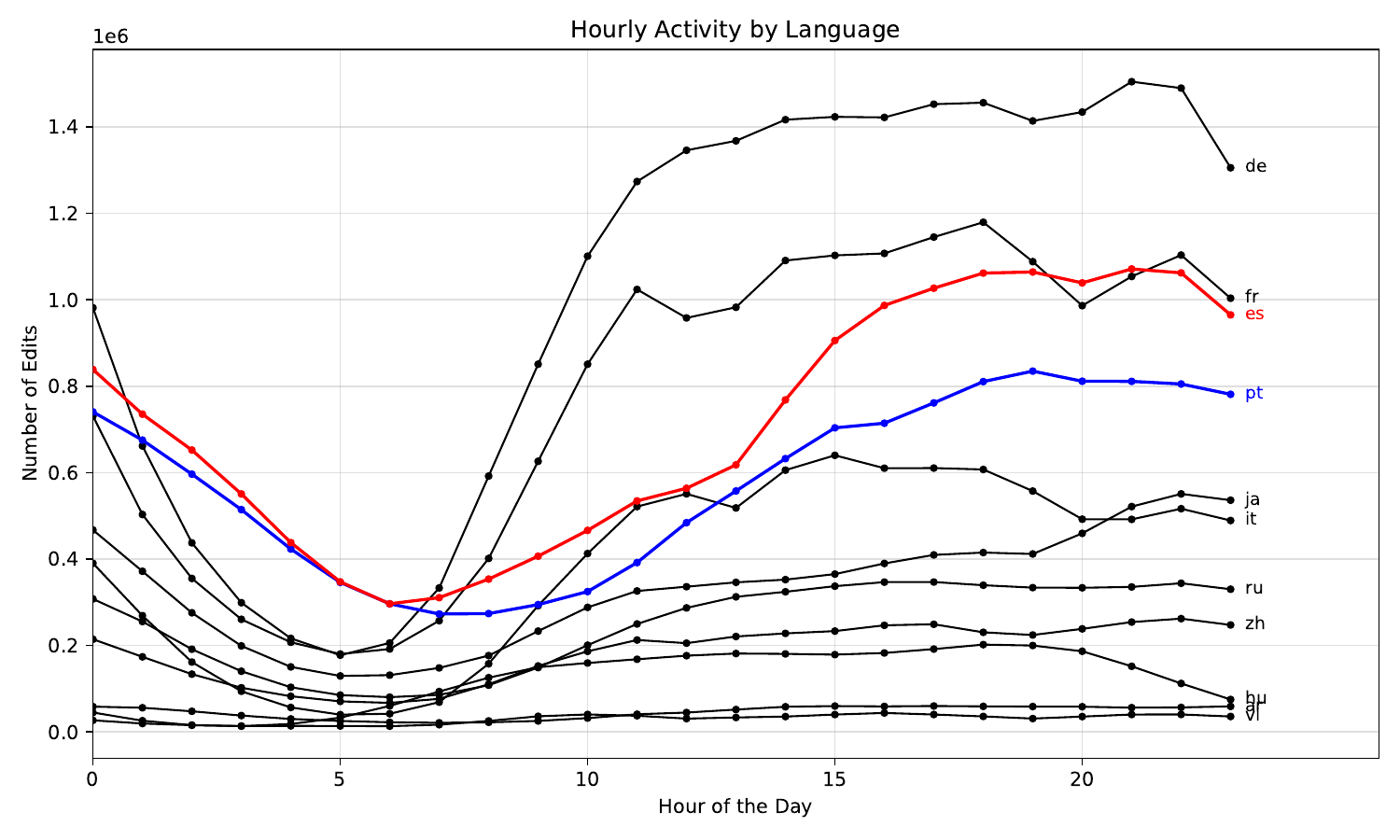}
    \caption{Number of edits per hour over the entire length of the data span for
all languages, with Spanish highlighted in red and Portuguese in blue for
comparison. {Since a page may belong to several categories, an edit
contributes to every category to which its page belongs, and therefore
appears in more than one of the curves of Figures 1 and 2. Here, where all
categories are aggregated into a single curve per language, each edit is
counted only once, so that the vertical axis is the number of edits per
hour.}
\label{fig3}}
\end{figure}

{Yasseri et al. \cite{Yasseri2012} reported that, whereas most language editions share a common circadian profile, the Spanish and Portuguese editions exhibit a slight delay in their daily activity, which they attributed to the time-zone offset introduced by the large volume of edits originating in Latin America. Figure \ref{fig3}, which aggregates the number of edits per hour of the day over the entire data span, reproduces this effect: the ES and PT curves (highlighted in red and blue) are visibly shifted relative to the remaining languages.}

{This shift raises a methodological concern specific to our temporal analysis. This is a difficult question to answer: while these two communities share many cultural traits, it is also true that a large community of former colonies sharing the same time zones could introduce a bias in the temporal clustering results, drawing the two editions together for reasons of shared time-zone structure rather than genuine cultural affinity. To gain insight into this issue, we turn to the static clustering (Section 3.2), which by construction contains no temporal information: it is built solely from the number of edits and editors per category. If ES and PT continue to cluster together once the temporal dimension is removed, their proximity is unlikely to be a mere artefact of the time shift, and instead points to a genuine similarity in editorial preferences.}

\subsection{Static clustering analysis} 
\label{sec:static}
\begin{figure}
\centering
\includegraphics[width=1\linewidth]{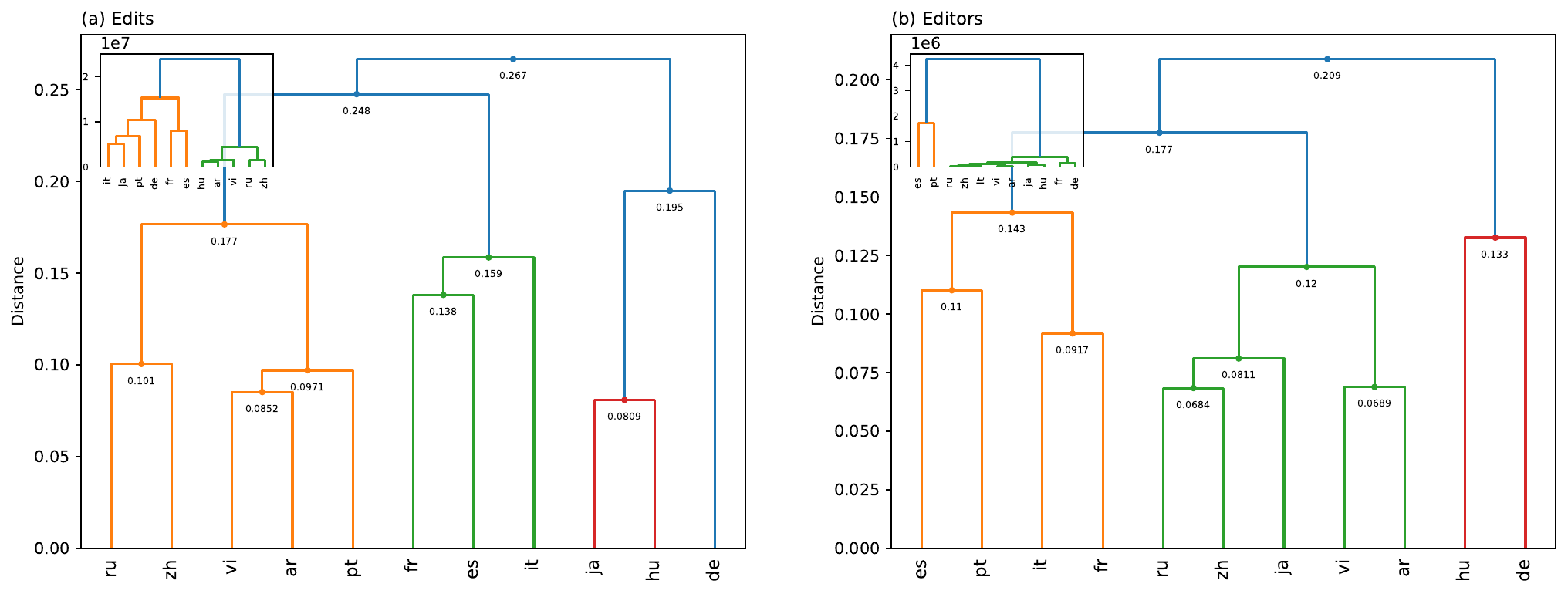}
\caption{Dendrograms of edits and editors for the static clustering of the dataset. On the left (a) the dendrograms of edits: {the normalised data} is shown prominently, with the non-{normalised} data in the inset. On the right (b), the dendrograms of editors: the {normalised} data is also shown in large format, with the non-{normalised} data displayed in the inset. {Normalisation divides each edition by its own total, so that its thirteen category values add up to one.}
\label{fig4}}
\end{figure}

{Figure \ref{fig4} presents the dendrograms for the number of edits (a) and the number of editors (b) by categories for the Wikipedia languages considered. Temporal patterns have not been considered at this stage: each language edition is described by the number of edits (or editors) accumulated in each of the thirteen categories over the whole period. The insets are built on these counts directly. In the main panels the counts of each edition are divided by their sum, so that the thirteen values add up to one and give the share of its edits (or editors) falling in each category. Editions of very different size therefore contribute equally, and the clustering responds only to how each community distributes its edits (or editors) across categories.}

{For the number of edits, two major clusters emerge in the non-normalised data: one comprising Western languages along with Japanese, and another with predominantly Eastern languages, including Hungarian. However, this division does not persist after normalisation, which suggests that it was driven by the disparity in size between the editions rather than by their thematic preferences. In the normalised data, a smaller cluster formed by Japanese, Hungarian and German separates from the remaining editions. In the second main cluster, some paired subgroups emerge, such as RU with ZH and AR with VI, the latter joined by PT; and ES with FR, joined by IT. Two of these pairs, ES with FR and RU with ZH, appear in both the normalised and the non-normalised data, indicating that their proximity follows from the categories these communities edit and not from the volume of their activity.}

When examining the dendrograms for the number of editors, the most striking observation is {the pair formed by ES and PT. In the inset, the two editions form a cluster completely separated from the rest. This could simply reflect the fact that these two editions have far more editors than the others, plausibly because of the large Spanish- and Portuguese-speaking populations outside Europe, in which case the pair would say more about size than about preferences. However, ES and PT remain a pair in the main panel, where each edition has been divided by its own total and size no longer plays any role, and where they join FR and IT before the remaining editions; their proximity therefore also reflects the categories their communities choose to edit. Moreover, since temporal patterns have not been considered at this stage, this grouping cannot be attributed to the time-zone shift discussed above: the proximity between ES and PT is therefore unlikely to be an artefact of their shared Latin American time zones, and instead points to a genuine similarity in editorial preferences. The other languages are mostly clustered in pairs, such as RU with ZH and AR with VI, both of which also appear in the analysis of edits, together with FR with IT and DE with HU.}

Comparing the two dendrograms (number of editors and number of edits) for both {normalised and non-normalised} versions, we observe greater consistency in the editor-based clustering. {Since the two versions differ only in the size of each edition, this indicates that it depends less on size}. We can also mention that the aggregation of RU and ZH is solid, appearing in every scenario considered. Possible reasons for this and the other proximities observed here are discussed in Sections \ref{dis} and \ref{con}.

\subsection{Temporal clustering analysis} 



\begin{figure}
    \centering
    \hspace{-0.5cm}
    \includegraphics[width=1.08\linewidth]{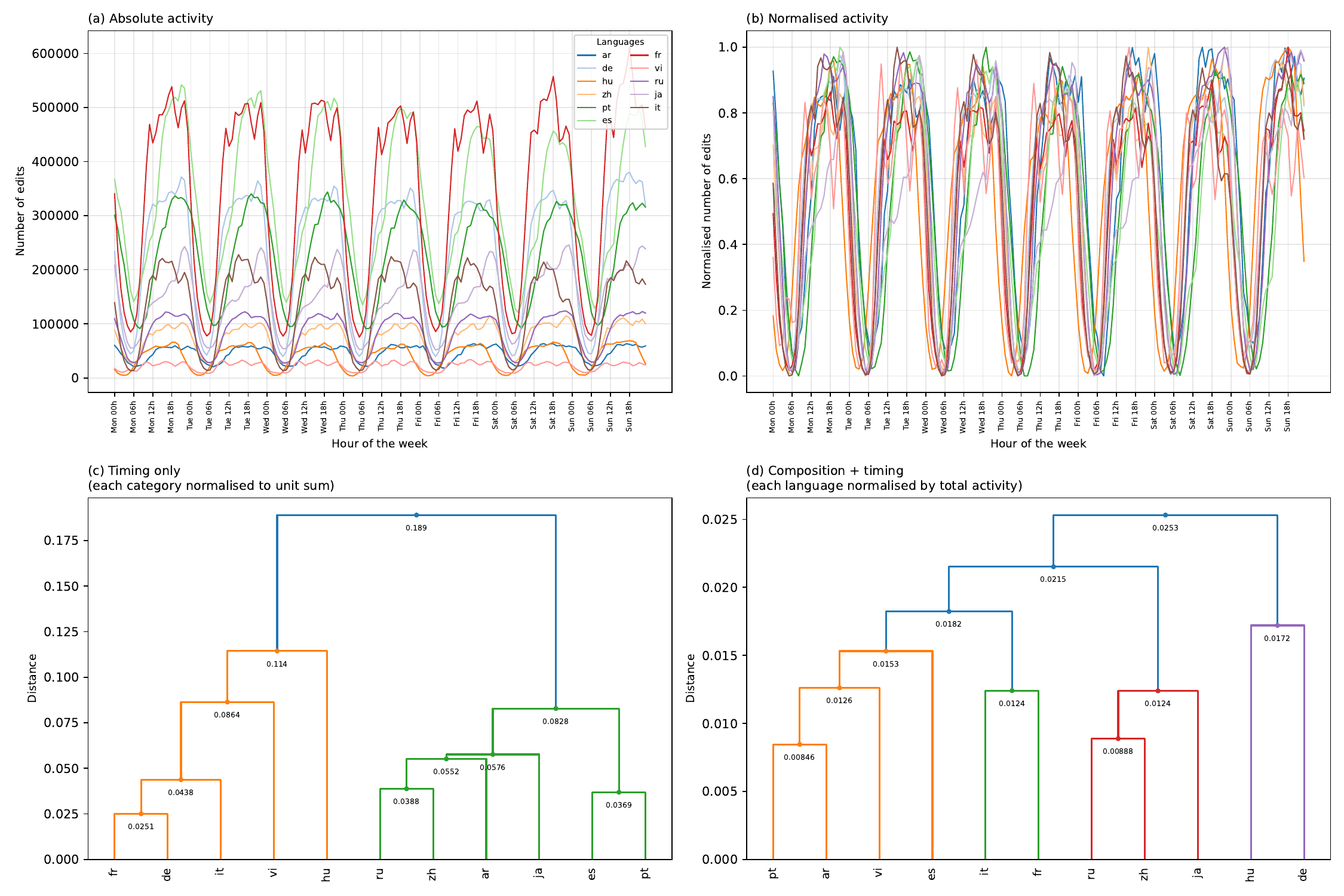}
    \caption{{Weekly edit activity patterns and hierarchical clustering. Top: absolute (a) and normalised (b) edit activity across different languages, where each language has been rescaled to the interval [0,1] so that the shape of its weekly cycle can be compared across editions of very different size. Bottom: dendrograms obtained from two complementary representations of the same data, both of which remove the effect of edition size. In (c), each category of each edition is normalised to unit sum over the 168 hours, so that the weight of the categories is removed as well and only the timing of the activity remains. In (d), each edition is divided by its total activity, so that the relative weight of the categories is preserved and the clustering reflects thematic composition together with timing.}
\label{fig5}}
\end{figure}

{In Figure \ref{fig5}, the upper part illustrates the weekly activity of each Wikipedia language. Panel (a) on the left shows the raw data, while panel (b) on the right displays the normalised one, each language rescaled to the interval [0,1] so that the shape of its weekly cycle can be compared across editions of very different size. The values were obtained for each hour of the week across the entire time period. The lower part presents the dendrograms obtained from two complementary representations of the same data, both of which remove the effect of edition size but differ in what else they remove. In panel (c), each category within each language is divided by its own total over the 168 hours, so that every category profile adds up to one; the relative weight of the categories is thereby removed as well, and the clustering responds only to timing, i.e., the shape of the weekly cycle. In panel (d), all the values of a language are instead divided by a single number, its total activity, so that they add up to one; because the divisor is the same for every category, the relative weight of each category is preserved, and the clustering responds to thematic composition together with timing. Consider two editions that both concentrate their sports activity on Sundays, but one of which devotes a fifth of its edits to sports while the other devotes a fiftieth: they appear close in panel (c), which sees only when the activity occurs, and far apart in panel (d), which also sees how much of it there is. The distinction matters because the category shares of Figure \ref{fig4} are recovered by summing the weekly profiles of panel (d), so that representation contains the static analysis within it, whereas panel (c) is by construction independent of it.}

Focusing first on the raw data, the most active Wikipedia languages are French (FR) and Spanish (ES). A second group of high-activity languages includes German (DE) and Portuguese (PT). The remaining languages follow, with Vietnamese (VI) showing the lowest activity levels.

{In terms of clustering, the two panels answer different questions. Panel (c) isolates the rhythm of editing. By normalising every category of every edition to unit sum, it removes both the size of the edition and the weight of each category, so that two communities appear close when they edit the same categories at the same hours of the week, regardless of how much they edit and of how prominent each theme is for them. It is therefore independent of the thematic composition that the static analysis of Figure \ref{fig4} measures. In this representation the root of the dendrogram separates two branches, one formed by FR, DE, IT, VI and HU, and the other by RU, ZH, AR, JA, ES and PT. FR and DE form the closest pair of the whole dendrogram, with IT attaching to them and VI and HU joining the group at successively greater heights. In the second branch, ES with PT and RU with ZH form pairs at almost identical heights, just above that of FR and DE; AR and JA attach to RU and ZH, and ES and PT join this group only afterwards. The proximity between ES and PT appears here both for edits and for editors, whereas in the static analysis the pair appears only for editors (Figure \ref{fig4}b). Panel (d), in contrast, combines what is edited with when it is edited: by dividing each edition by a single number, its total activity, it retains the relative weight of the categories and therefore combines thematic composition with timing. In this representation the root separates HU and DE from the remaining editions, which in turn divide into a group formed by RU, ZH and JA and another formed by PT, AR, VI, ES, FR and IT. PT and AR form the closest pair, with VI and then ES attaching to them, while RU pairs with ZH, FR with IT and DE with HU. The broad division resembles that of the static analysis, where DE and HU also separate early from the rest, but at the level of pairs only RU–ZH is common to both, as discussed below in relation to Figure \ref{fig7}.}

\begin{figure}
    \centering
    \hspace{-0.5cm}
    \includegraphics[width=1.08\linewidth]{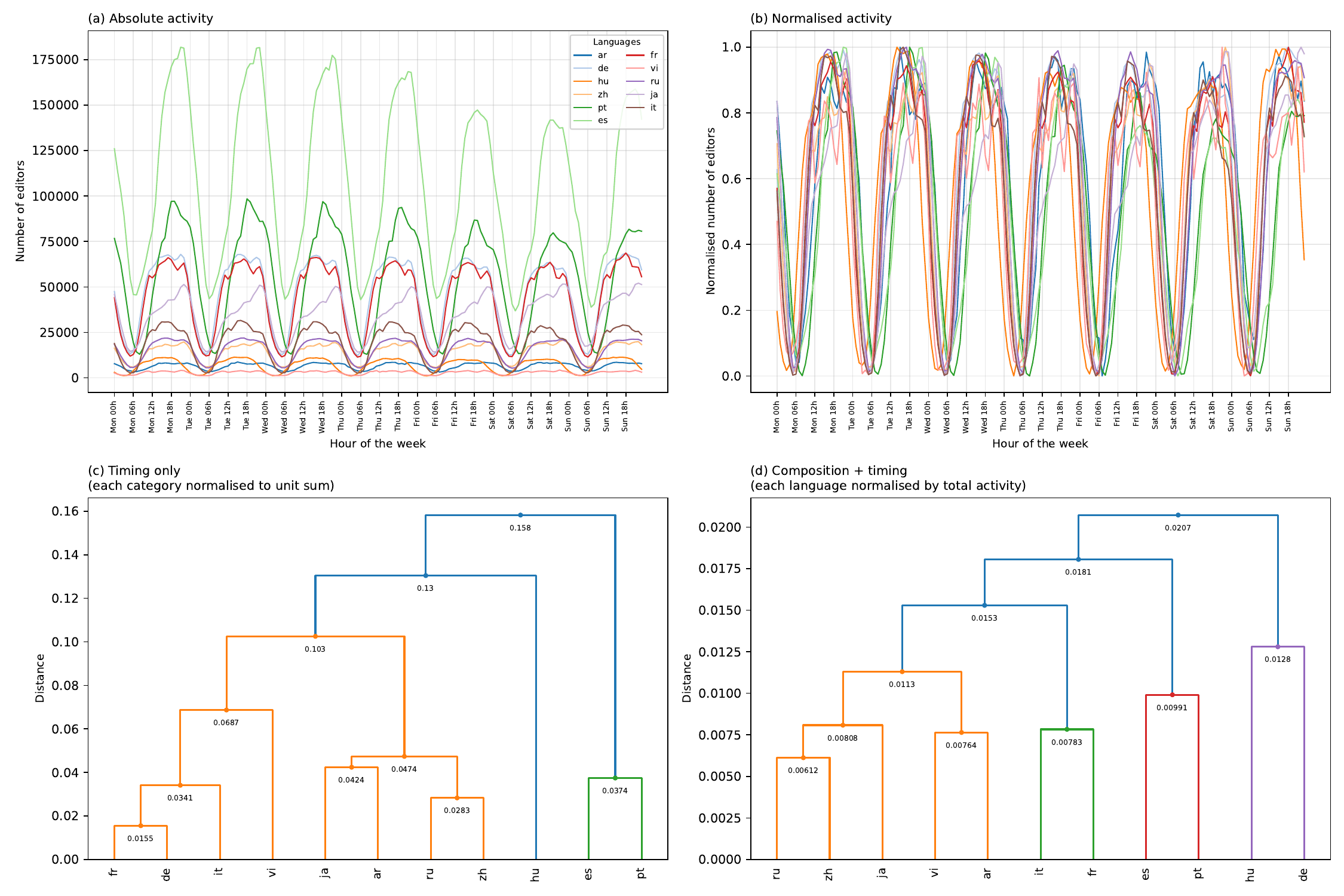}
    \caption{Weekly editor activity patterns and hierarchical clustering. Top: absolute (a) and normalised (b) editor activity across different languages, where each language has been rescaled to the interval [0,1] so that the shape of its weekly cycle can be compared across editions of very different size. Bottom: dendrograms obtained from two complementary representations of the same data, both of which remove the effect of edition size. In (c), each category of each edition is normalised to unit sum over the 168 hours, so that the weight of the categories is removed as well and only the timing of the activity remains. In (d), each edition is divided by its total activity, so that the relative weight of the categories is preserved and the clustering reflects thematic composition together with timing.
\label{fig6}}
\end{figure}

\begin{figure}
    \centering
    \hspace{-0.5cm}
    \includegraphics[width=1.08\linewidth]{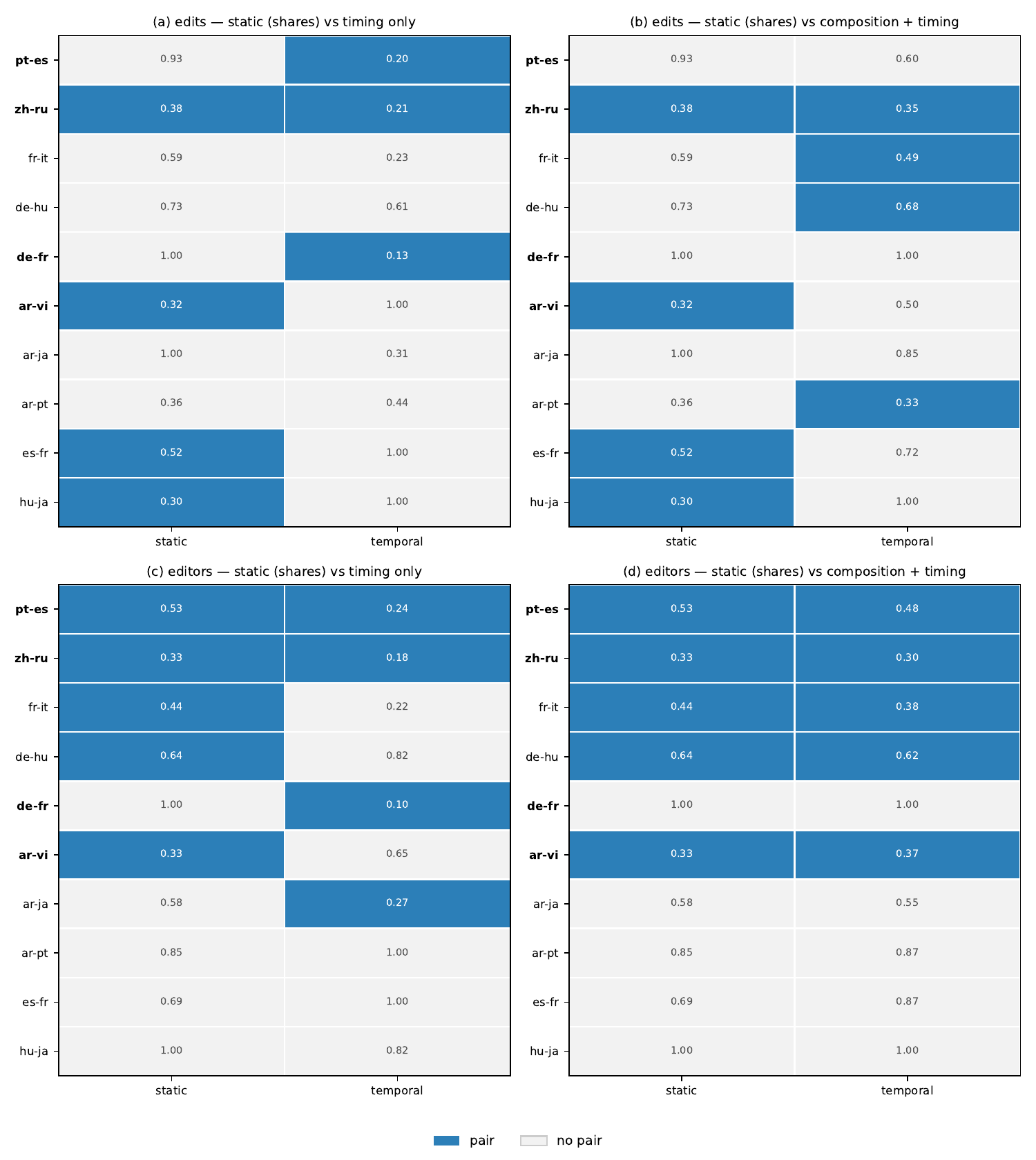}
    \caption{Language pairs in the static and in the temporal clustering. Rows are the pairs that form in at least one of the analyses; a cell is filled when the two editions join each other directly in that analysis, and the number is the height at which they do so relative to the root of the dendrogram, so that 0 means they join immediately and 1 that they meet only at the root. Panels (a) and (c) are the comparison of interest: they contrast the category shares of Fig. 4 with the timing-only representation of Figs. 5c and 6c, which is independent of thematic composition by construction, so that any difference between the two columns can only come from the temporal dimension. Panels (b) and (d) serve as a reference: the category shares are recovered by summing the weekly profiles of the composition-and-timing representation, and composition accounts for most of the distance in it, so the two columns are expected to agree closely. They need not agree exactly, since clustering the 2,184-dimensional representation is not the same as clustering its 13-dimensional marginal: the timing contributes the remaining part of the distance and can reorder fusions that lie at similar heights.
\label{fig7}}
\end{figure}

In Figure \ref{fig6}, we present the weekly activity for the number of editors, following the same format as Figure \ref{fig5}. {The upper section displays the raw data (panel a) on the left and the normalised data (panel b) on the right. The lower section features the dendrograms obtained from the two representations described above: timing only (panel c) and thematic composition together with timing (panel d).} A striking observation is the significantly higher activity of Spanish (ES) editors compared to the others, followed by Portuguese (PT) in second place, with German (DE) and French (FR) ranking third and fourth, respectively.

{Plotting all editions on common axes makes visible patterns that the individual panels of Figures \ref{fig1} and \ref{fig2} do not reveal, since there each edition is shown on its own scale. The German and French editions display almost identical weekly profiles, with their daily maxima and minima occurring at the same hours throughout the week and converging entirely at weekends, while the Italian edition follows the same shape at a lower level of activity. This is reflected in the clustering of panel (c), where FR and DE form the closest pair and IT attaches to them at more than twice that distance. The German edition further distributes its weekday activity over more than one daily maximum, with the central one clearly more pronounced, and concentrates it into a single peak at around 6 pm at weekends. The Spanish edition, for its part, shifts its daily maximum from 8 pm on weekdays to 6 pm at weekends, a displacement that no other edition presents.}

{In terms of clustering analysis, ES and PT form a pair in both representations, as they did in the static analysis of editors. In panel (c), which retains only the timing of the activity, they separate from all the remaining editions at the root of the dendrogram. FR and DE form the closest pair of that dendrogram, with IT attaching to them at more than twice that distance. RU and ZH also group together, and they do so in both representations, as they did in the static analysis. The remaining groupings differ between the two panels: in panel (c) VI joins the group formed by FR, DE and IT, and AR pairs with JA, whereas in panel (d), where thematic composition is retained as well, FR pairs with IT instead of DE, VI pairs with AR, and DE pairs with HU.}

{Figure \ref{fig7} contrasts the pairs that form in the static clustering with those that form in each of the two temporal representations. The two temporal representations play different roles in this comparison. The timing-only representation carries no information about thematic composition, so that a pair that forms in it but not in the static analysis can be attributed to shared rhythm alone; the composition-and-timing representation, in which composition accounts for most of the distance, serves as a reference. For the number of editors, of the five pairs present in the static analysis only two, ES with PT and RU with ZH, persist once the weight of the categories is removed; FR with IT, DE with HU and AR with VI disappear, and two pairs emerge that the static analysis does not show, DE with FR and AR with JA. The case of DE and FR is the clearest: the two editions do not group at all in the static analysis, where they meet only at the root of the dendrogram, and yet they form the closest pair of the timing-only one. Their communities edit different categories, but they edit them at the same hours. The composition-and-timing representation behaves differently: for editors it recovers all five static pairs, which confirms that the divergence just described follows from removing the relative weight of the categories and not from the change in dimensionality from 13 to 2184 coordinates. For the number of edits the agreement with the static analysis is weaker in both representations, with only RU and ZH common to them, which is consistent with the static pairs of the edit-based analysis being less stable than those of the editor-based one, as already noted in Section 3.2. Possible reasons for the DE–FR and RU–ZH proximities are discussed in Sections \ref{dis} and \ref{con}.}

\section{Discussion}\label{dis}

In this research, we have seen how Wikipedia data can show robust patterns of digital activity specific to each of the languages studied. Even though our analysis is limited to Wikipedia editing, {previous studies have shown that Wikipedia activity is a useful proxy for the interests of each linguistic community} \cite{Miccio2025,Gandica2018,Samoilenko2016,Hecht2010}.

The study was carried out by separating the different categories into the main tree-branches defined in the English Wikipedia and maintained by its volunteer editor community. We first analysed the temporal evolution for both the number of editors and edits, averaging the length of the data (approximately the first ten years of Wikipedia) in each of the hours of a full week. This dual-layered approach (edits and editors) mitigates the potential bias introduced by individual saving patterns or editorial conflicts: {when two distinct activity metrics converge on the same grouping, this convergence is unlikely to arise by chance,} thus strengthening the stability and robustness of the findings.

{The most edited categories are history for HU (Hungarian), ES (Spanish), FR (French), PT (Portuguese) and JA (Japanese), art for DE (German), and nature for RU (Russian) and ZH (Chinese); the preference for history in HU and JA, for art in DE and for nature in RU was already reported in \cite{Gandica2018}.} Here we can see some other nuanced behaviours, for example, VI (Vietnamese) has three daily peaks of activity contrasting with the other languages. The Japanese, Italian, and Portuguese Wikipedias exhibit peaks of editing activity mostly at 21h, 15h and {18h}, respectively. ES, PT (Portuguese) and IT (Italian) are more active on weekdays, while FR and ZH (Chinese) are more active on weekends. Several other interesting patterns can be seen if looking carefully at Figures \ref{fig1} and \ref{fig2}. For example, we were able to confirm the increased activity towards non-working hours (evenings and weekends), as mentioned by Yasseri et al. \cite{Yasseri2012} for Asian and Arabic languages, with the particularity of Friday also being an active day of the weekend in the AR-WP (Arabic) because it is a public holiday, as also mentioned in \cite{Yasseri2012}.

Regarding the static clustering, we found the editor-based clustering to be
more robust than the edit-based clustering, in terms of the results for
{normalised and non-normalised} data. We can highlight a marked closeness
between ES and PT {in the number of editors}. The same pair presents a
particular shift when looking at the average daily number of edits, which it
is natural to attribute to the contributions that both languages must have
from Latin American countries (Figure \ref{fig3}). However, since the static clustering contains no temporal information, their closeness in terms of edited categories is unlikely to be an artefact of this shift.

{Moreover, the comparison between the static and the temporal clusterings
shows that some proximities depend on which aspect of activity is retained.
German (DE) and French (FR) do not group in the static analysis, where they
meet only at the root of the dendrogram, yet they form the closest pair of
the timing-only representation, for both edits and editors: their
communities devote different shares of their activity to each category, but
they edit each of them at the same hours. Vietnamese (VI) and Arabic (AR)
show the opposite behaviour: they group in the static analysis and, for
editors, when composition is retained together with timing, but not on the
basis of timing alone, so that their proximity rests on the thematic content
they edit rather than on shared rhythms of activity. Temporal dynamics
therefore offer a distinct lens on cultural proximity, one that separates
similarities of rhythm from similarities of content. Furthermore, as with
the convergence noted above, Russian (RU) and Chinese (ZH) group together in
the static analysis and in both temporal representations, for edits and for
editors, which suggests a robust pattern that holds across representations and activity metrics. One possible explanation is the length of working hours, which, as noted by Yasseri et al.~\cite{Yasseri2012}, tends to be longer in Russia and China~\cite{ilo}.}

Our analysis reveals both shared and divergent temporal editing patterns
across the examined Wikipedia language communities. {Beyond the global
circadian trends common to all editions, each community shows distinctive
deviations that appear to be shaped by cultural and social factors. Whereas
Technology Appropriation Theory accounts for the divergence in thematic
preference documented above, these temporal deviations call for a
complementary framework, which we find in Structuration Theory ~\cite{Giddens2014}. This
theory posits that social practices, such as Wikipedia editing, are produced
and reproduced through the dynamic interplay between individuals and broader
social structures, including cultural norms, routines, and institutional
contexts.} In this framework, the observed editing patterns are not merely the
result of individual preferences, but also reflect the ongoing interaction
and transformation of collective temporal and cultural structures within each
linguistic community. These insights not only deepen our understanding of
digital cultural behaviour but also open avenues for future research into
the temporal dynamics of online collaboration.

\section{Conclusions}
\label{con}
This study explored the temporal {patterns of thematic preferences in the
editing activity of Wikipedia communities} across 11 languages, uncovering
insights into cultural dynamics and digital behaviour. The conclusions of this work span three main dimensions: data insights, methodological contributions, and cultural implications.

By analysing both the number of edits and the number of editors as proxies
for user activity, we find that editor-based metrics provide the more
consistent measure: the editor-based clusterings agree more closely across
normalisations, and between the static and the temporal analyses, than the
edit-based ones. Counting editors rather than edits also limits the
influence of individual saving patterns and of editorial conflicts, such as
edit wars, in which the same content is repeatedly modified by opposing
editors.

{From a methodological standpoint, the study brings together two components of collective editing activity that previous studies have examined only in isolation: the
thematic composition of each community and the timing with which it edits
each theme. Normalising every category to unit sum isolates the latter, and
its comparison with a representation that retains composition shows which
proximities rest on shared rhythms and which on shared content. A second
methodological point concerns dimensionality reduction, which is not a
neutral step for this kind of question. If two communities edit the same
category at the same hours of the week, that similarity is confined to a few
of the coordinates of the representation; unless those coordinates happen to
align with the directions of greatest variance across editions, a reduction
such as principal component analysis discards them, and the clustering no
longer answers the question it was meant to address. Analogous concerns
apply, in different forms, to nonlinear methods. For this reason the main
analyses are carried out on the full representation. When the data are
nonetheless reduced before clustering, as reported in the Supplementary
Information, most of the groupings survive, but how many depends on the method chosen, which confirms that the reduction is not neutral.}

Turning now to the results of our study, we note that, while previous studies have conducted a static study of preferences within categories or coarse-grained studies of temporal activity in different languages, our study performs a fine temporal analysis of activity in different categories across different languages, and more importantly, we perform a clustering-based study to find temporal closeness between different language
populations around the globe. In this way, we have found unexpected and noteworthy patterns, such that the French edition of Wikipedia does not group with the German one in a static analysis, while {it is the closest to it when only the timing of the activity is taken into account: the two communities devote different shares of their activity to each category, but edit each of them at the same hours. This proximity becomes visible only
when every category is normalised separately; when each edition is normalised as a whole, the differences in thematic composition dominate the distance and mask the shared rhythm. The closeness of the Chinese and Russian language communities behaves in the opposite way: it does not depend on this choice, since it appears in the static analysis and in both temporal representations, for edits and for editors alike, and would therefore be
found under any of them.} Those unexpected cultural proximities merit further
exploration.

{The Portuguese and Spanish languages were consistently grouped in all analyses of the number of editors, static and temporal. Since the static analysis contains no temporal information, their proximity in preferences does not seem to be an effect of the time shift due to both having a highly active population in the American continent, but rather reflects the preferences of these two communities as a whole, on both continents.}

{Despite the robustness of our findings, it is important to consider the study’s limitations. The first is to assume that the digital activity is representative of the social preferences of the entire community. In this sense, we can consider that our society is increasingly immersed in the digital world and that the percentage of society that is left out of this dimension is becoming smaller and smaller. Then, the best we can do as computational social scientists is to refine the ways in which these resources can be useful to analyse the emergence of the different dimensions manifested by our societies.}

The second limitation concerns the effect of having aggregated the editors' activity. Although different digital media (data from calls, sms and e-mails) have shown that there exist individual persistent particularities on daily cycles \cite{Aledavood2015}. Here, we have seen that the patterns of maximum and minimum activity at the system-level averages are quite clear and specific to each language population.

{The third limitation is of a technical nature. Specifically, the fact that we have removed bot-generated edits, excluding all edits with the word “bot”, in any combination of upper or lower case letters, of usernames. It would be interesting to perform the data-cleaning using the Wikipedia Users API end-point. However, only official bots would be identified when using the API. Many editors use automatic programs to correct editing problems without flagging them as bots. That issue of bots is generally complicated by analysing big data and Wikipedia data is not exempt from this.}

The last limitation is rather a possible extension to the present work, it has to do with the fact that we have analysed the people who edit Wikipedia and not those who read it. In the spirit of the work done by {TeBlunthuis et al. \cite{TeBlunthuis2019} and} Piccardi $\&$ West et al. \cite{Piccardi2024}, it would be very interesting to carry out an analysis like the one done here but for the other side of the coin, i.e., to analyse the Wikipedia reading population. In the present work, we consider the efforts made by Wikipedia’s volunteer editors: confronting conflicts, combating vandalism, and participating in the demanding editing process as a proxy to calibrate the interests of their respective communities. These editorial actions are driven by a genuine desire to preserve cultural memory, portray reality through their own lens, and document their people, places, and events. { A further extension concerns the range of languages: the eleven editions analysed here were inherited from \cite{Gandica2018}, where they were selected as a trade-off
between global coverage and the size of the corresponding Wikipedia
editions, rather than under an explicit sampling criterion. Extending the
analysis to a substantially larger and systematically selected set of
editions would allow the patterns reported here to be tested for generality,
and would make it possible to relate temporal editing patterns to
article-level outcomes such as content quality.}

These insights extend beyond Wikipedia to other digital platforms, offering strategies to optimise content management and resource allocation. From a practical standpoint, the findings carry direct implications for Wikipedia's governance and community management. Knowing when and in which categories each linguistic community is most active could help Wikipedia administrators and the Wikimedia Foundation design targeted anti-vandalism patrolling strategies, schedule editorial campaigns, and allocate review resources aligned with the cultural rhythms of each community. For instance, communities with pronounced weekday activity peaks, such as ES and PT, may benefit from different moderation schedules than those with weekend-concentrated editing, such as FR and ZH.

From a theoretical perspective, the divergence between static and temporal
clustering results raises important questions about the cultural factors
underlying cross-linguistic proximities. {The clearest case is that of DE
and FR, whose editions do not group in the static analysis but form the
closest pair once only the timing of the activity is retained. Samoilenko et
al. \cite{Samoilenko2016} and Yoon et al. \cite{Yoon2023} account for
cross-linguistic proximities in Wikipedia through mechanisms grounded in
interaction between communities: the former identify linguistic similarity,
shared religion, and demographic attraction as predictors of multilingual
co-editing activity, while the latter attribute knowledge-structure
similarity primarily to socio-economic ties over geographical proximity.
Neighbouring communities with strong socio-economic ties, such as the German
and French ones, are precisely those that these mechanisms would lead us to
expect close in what they edit; yet the two editions are distant in content
and close only in rhythm. This suggests that structural similarities in
daily routines, such as work–leisure transitions or evening engagement
patterns, may constitute a dimension of cultural affinity distinct from the
one captured by interaction-based measures. Detecting a proximity of this kind requires a thematic partition of the activity shared by the communities compared; traces that record when people act, but not what they act upon, can show that cultures differ in their rhythms, but not that two cultures share a rhythm while differing in content.}

{The RU--ZH pair behaves
differently: it appears in the static analysis and in both temporal
representations, so that its proximity is compatible with interaction-based
mechanisms as well as with shared structural features of daily life, such
as longer working hours \cite{ilo}, which we consider a plausible candidate
for their convergence. This reading must be qualified by the fact that the
Chinese edition has been blocked in mainland China since 2015, outside our
period of analysis, with intermittent restrictions dating back to
2004 \cite{wikimedia2019block}, so that part of its activity may originate
outside mainland China. Taken together, these cases illustrate a distinction
between the temporal rhythm of preferences and preferences in absolute terms
that our approach makes visible. These observations invite future research
combining Wikipedia editing data with sociodemographic indicators to
disentangle the mechanisms behind these cultural proximities.}

{Moreover, in
an era where AI-generated digital encyclopedias are beginning to emerge, the
temporal editing patterns captured in platforms like Wikipedia acquire
additional value as an authentic record of the cultural preferences and
rhythms of linguistic communities, a type of behavioural information that
automated content platforms cannot provide.}


{In summary, this work shows that Wikipedia editing data can be used to compare collective behaviour across linguistic communities, and that the timing of that behaviour reveals structure that thematic composition alone does not. Time is therefore a relevant, and often neglected, dimension in cross-cultural studies of digital trace data.}

\section*{Codes and Data availability}
All codes used are available at https://github.com/yerali/wikipedia-temporal-patterns. The data are available at https://zenodo.org/records/15807858, with doi: https://doi.org/10.5281/zenodo.15807858.

\section*{Clinical trial number} not applicable

\section*{Competing interests} The authors declare that they have no competing interests as defined by Springer, or other interests that might be perceived to influence the results and/or discussion reported in this paper. 

\section*{Acknowledgements}
YG thanks Julieta Barba for valuable discussions related to physiological and socially induced circadian patterns.


\section*{Authors' contributions} YG  designed the research, DV performed the calculations, YG wrote the manuscript. All authors read and approved the ﬁnal manuscript.

\bibliography{sn-bibliography}

\end{document}